\documentclass[usegraphicx,usenatbib]{mn2e}

\newcommand{\xmm}{{\it XMM-Newton}}

\newcommand{\exosat}{{\it EXOSAT}}
\newcommand{\rosat}{{\it ROSAT}}

\newcommand{\euve}{{\it EUVE}}

\title{XMM-Newton observes flaring in the polar UZ For during a low state}

\author[Dirk Pandel and France A. C\'ordova]
{Dirk Pandel and France A. C\'ordova\\
Department of Physics, University of California, Santa Barbara, CA 93106, USA}

\pagerange{\pageref{firstpage}--\pageref{lastpage}}
\pubyear{2002}

\begin{document}

\maketitle

\label{firstpage}


\begin{abstract}

During an \xmm\ observation, the eclipsing polar UZ For was found in a peculiar 
state with an extremely low X-ray luminosity and occasional X-ray and UV 
flaring.
For most of the observation, UZ For was only barely detected in X-rays and 
$\sim\!800$ times fainter than during a high state previously observed with 
\rosat.
A transient event, which lasted $\sim\!900\:$s, was detected simultaneously by 
the X-ray instruments and, in the UV, by the Optical Monitor.
The transient was likely caused by the impact of $10^{17}$--$10^{18}\:$g of
gas on the main accretion region of the white dwarf.
The X-ray spectrum of the transient is consistent with $\sim\!7\:$keV 
thermal bremsstrahlung from the shock-heated gas in the accretion column.
A soft blackbody component due to reprocessing of X-rays in the white dwarf 
atmosphere is not seen.
The increase in the UV flux during the transient was likely caused by cyclotron 
radiation from the shock-heated gas.
Two more flaring events were detected by the Optical Monitor while the X-ray 
instruments were not operating.
We conclude from our analysis that the unusual flaring behavior during the low 
state of UZ For was caused by intermittent increases of the mass transfer rate 
due to stellar activity on the secondary.
In addition to the transient events, the Optical Monitor detected a roughly 
constant UV flux consistent with 11000--K blackbody radiation from the 
photosphere of the white dwarf.
We find a small orbital modulation of the UV flux caused by a large,
heated pole cap around the main accretion region.

\end{abstract}

\begin{keywords}
Stars: individual: UZ For --
Stars: novae, cataclysmic variables --
Stars: binaries: eclipsing --
Stars: magnetic field --
Stars: activity --
X-rays: binaries
\end{keywords}


\section{Introduction}

UZ For is an eclipsing member of the subclass of cataclysmic variables
called AM Her binaries or polars.
In these binaries, the strong magnetic field of the white dwarf primary causes 
it to rotate synchronously with the orbital motion.
The magnetic field also prevents the formation of an accretion
disk around the white dwarf.
The accretion stream from the Roche-lobe filling secondary
is funneled along the magnetic field lines and impacts the
white dwarf near a magnetic pole.
Slightly above the surface, the accretion stream forms
a stand-off shock that heats the gas to temperatures
in excess of $10^8\:$K.
The shock-heated plasma then cools and settles on to the white dwarf
while strongly emitting cyclotron radiation (IR to UV)
and thermal bremsstrahlung (mostly X-rays).
The photosphere below the shock is heated by reprocessing of X-rays
and emits blackbody radiation visible at soft X-ray and UV energies.
Many polars have been observed to decline in brightness by several 
magnitudes and remain in a faint state for days to years.
The causes of these low states are not known, but, in the absence of an 
accretion disk, the large brightness variations must be due to changes in 
the mass transfer rate from the companion star.
A comprehensive review of polars is given in \citet{1995cvs..book.....W}.

UZ For (EXO 033319-2554.2) was first detected as a serendipitous X-ray
source with \exosat\ \citep{1987IAUC.4486....1G,1988ApJ...328L..45O}.
Subsequent optical spectroscopy and polarimetry established UZ For as an
eclipsing polar \citep{1988A&A...195L..15B,1988ApJ...329L..97B}.
The 126.5--min orbital period is close to the lower edge of the 2--3 hr
"period gap", a sparsely populated region in the orbital period distribution of
cataclysmic variables.
UZ For is a high inclination system ($i\approx80^\circ$),
and both accretion regions are eclipsed by the white dwarf for at least half
of an orbital cycle.
The optical spectrum of UZ For shows strong cyclotron emission lines
which indicate magnetic fields of $\sim\!53\:$MG and $\sim\!48\:$MG for the two
accretion poles \citep{1996A&A...310..526R}.
The best available estimates for the distance and the white dwarf mass are
$d=208\pm40\:$pc and $M_{WD}\approx0.6$--$0.8\:M_\odot$
\citep{1989ApJ...337..832F,1991MNRAS.253...27B}.

In this paper we present X-ray and UV data obtained with \xmm\ 
while UZ For was in a state of unusually low and irregular accretion.
We study the properties of an X-ray/UV flare and show that it was
caused by accretion on to the white dwarf.
We argue that the intermittent accretion rate increase was due to 
stellar activity on the companion star.
Part of the \xmm\ data has recently been published in 
\citet{2001ApJ...562L..71S}. 
The goal of our paper is to utilize all available \xmm\ data and present 
additional results not previously published.
In particular, we find two more UV flares and clearly identify 
eclipses in the UV light curves.
We show that the beginning of the X-ray/UV flare coincides with the eclipse 
egress of the main accretion region.
This provides strong evidence that the flare was an accretion event on 
the white dwarf.
We estimate the total accreted mass and show that it is consistent with 
the mass ejected by stellar flares on M dwarfs.
We also detect a weak orbital modulation of the X-ray and UV fluxes and 
demonstrate that it is most likely due to emission from the main accretion 
region.


\section{Observations and analysis}

\begin{figure*}
\includegraphics{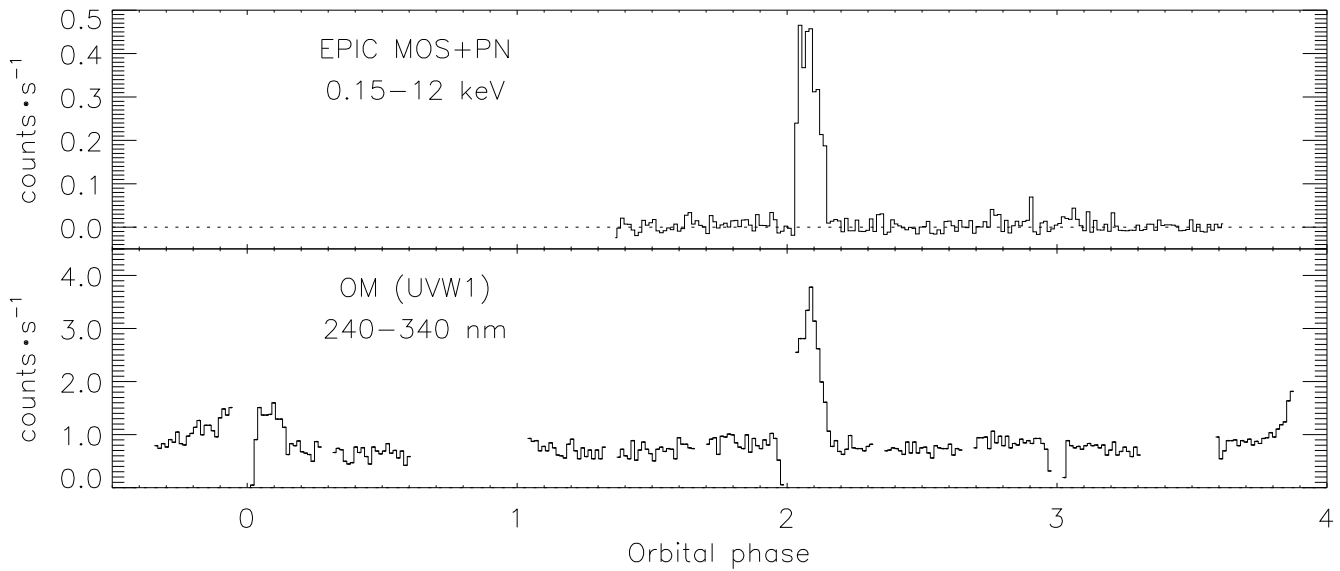}
\caption[X-ray and UV light curves]
{X-ray and UV light curves of UZ For obtained with \xmm\ on Jan 14, 2001
(binning 100 s).
Shown are the combined X-ray count rates of the three EPIC detectors
(upper panel)
and the UV count rate measured with the Optical Monitor using the UVW1 filter
(lower panel).
Orbital phase 0 is defined as the centre of the first eclipse as 
predicted by the ephemeris in \citet{2001MNRAS.324..899P}.
}
\label{lightcurve}
\end{figure*}

UZ For was observed with \xmm\ on 14 January 2001 as part of the
performance verification of the mission.
\xmm\ \citep{2001A&A...365L...1J} carried out two separate observations,
but during the first observation the EPIC cameras
\citep{2001A&A...365L..18S,2001A&A...365L..27T} were used for calibration
purposes, and only the Optical Monitor \citep{2001A&A...365L..36M} was
observing the source.
During the second observation each of the three EPIC instruments performed one 
long exposure (16 ks for PN; 18 ks for MOS) with the thin 
blocking filters and the CCDs in large window mode.
The Optical Monitor (OM) performed a total of 11 exposures of   
2200 s duration each (3 during the first and 8 during the second 
observation). All OM exposures were done in fast mode (0.25 s timing 
resolution) and with the UVW1 filter (240--340 nm).

From the EPIC MOS and PN data we extracted good photon events (FLAG=0)
using a circular aperture centred on UZ For and with a 20'' radius.
Count rates quoted throughout the paper have been corrected for the 75 per cent
enclosed energy fraction of this aperture.
The background rate was estimated from a larger region on the same CCD
as the source image.
We included in our analysis events in the energy range 0.10--12 keV for the 
MOS and 0.15--12 keV for the PN.
To maximize the signal-to-noise ratio, we used photon events
with patterns 0--12.
Only for the spectral analysis of the PN data did we select exclusively events 
with pattern 0.
To create X-ray light curves, we applied a barycentric correction to the photon 
arrival times and combined the events from all three EPIC instruments.
For the extraction of UV light curves from the OM fast mode data,
we used a circular aperture of 5.5'' radius,
which encloses 78 per cent of the detected source photons.
Background subtracted UV and X-ray light curves of the available data are shown 
in Fig. \ref{lightcurve}.
We defined as orbital phase zero the centre of the first eclipse as predicted by
the most recent ephemeris \citep{2001MNRAS.324..899P}, which has an accumulated
uncertainty of 4 s.


\subsection{X-ray light curve}
\label{xraylc}

Except during a short flare, the X-ray flux from UZ For (Fig. \ref{lightcurve})
was at an extremely low level never before seen in this object.
Even with the EPIC data integrated over the entire observation 
(excluding the flare), UZ For was detected only at a $4\:\sigma$ level.
An improved $6\:\sigma$ detection could be achieved by limiting the integration 
to the orbital phase range 0.65--0.15.
For these orbital phases, the main accretion region is on 
the side of the white dwarf facing us \citep{2001MNRAS.324..899P}.
For orbital phases 0.15--0.65, for which the main accretion region is not 
visible, the integrated source count rate is consistent with zero at a 
$1\:\sigma$ level.
It appears that the main accretion region was the source of the weak X-ray 
emission.
We cannot rule out the possibility that the X-rays originated from the
secondary region, which is visible for orbital phases 0.75--0.35.
Yet one might expect that, as for other polars during low states,
accretion was limited to the main region
\citep[e.g.][]{1989A&A...223..179B}.

The combined EPIC MOS+PN count rate, averaged over the orbital phase range 
0.65--0.15 (excluding the flare), is $0.009(2)\:$s$^{-1}$ 
(0.15--12 keV).
\citet{1993MNRAS.262..993R} found, based on \rosat\ observations, that the 
X-ray spectrum of UZ For in the high state is best described by an absorbed 
28 eV blackbody below 0.5 keV plus a weak thermal bremsstrahlung tail 
visible at higher energies.
Assuming this high-state spectrum, we would predict a \xmm\ count rate of 
$\sim\!7\:$s$^{-1}$, which is $\sim\!800$ times higher than observed. 
The EPIC spectrum does not show any evidence of the blackbody component that was 
seen with \rosat.
However, the spectrum is in good agreement with a thermal bremsstrahlung 
model.
It is possible that, due to a lower temperature of the accretion spot,
the blackbody component was shifted below 0.1 keV
and not detectable by the EPIC instruments.
This would require a blackbody temperature $\le\!15\:$eV,
provided that the emitting area was the same as during the \rosat\ observation.
If we consider only the weak bremsstrahlung component found in the \rosat\ 
spectrum, we would predict a \xmm\ count rate of $\sim\!0.26\:$s$^{-1}$.
This is still $\sim\!30$ times higher than observed.

We estimated the total bremsstrahlung flux (excluding the flare) by fitting a 
single temperature thermal bremsstrahlung model to the EPIC spectra.
With the temperature fixed at 10 keV, we obtained a bolometric bremsstrahlung 
flux $F_{brems}\approx2\times10^{-14}\:\mathrm{erg\:cm^{-2}\:s^{-1}}$.
Of this flux, 80 per cent was inside the 0.10--12 keV energy range and 
directly detected by \xmm.
For a distance $d=200\:$pc, the corresponding bolometric luminosity is
$L_{brems}\approx7\times10^{28}\:\mathrm{erg\:s^{-1}}$.
Here we used a geometric factor of $3.1\:\pi$, which takes into account the 
scattering of X-rays off the white dwarf's surface \citep{1987MNRAS.227..205K}.
Assuming that all of the accretion energy was emitted in the bremsstrahlung 
component, we can estimate the accretion rate $\dot{M}$ using
$L_{brems}\approx G\:M_{WD}\:\dot{M}/R_{WD}$.
For a white dwarf with a mass $M_{WD}=0.7\:M_\odot$ and a radius 
$R_{WD}=8\times10^8\:$cm, we find 
$\dot{M}\approx6\times10^{11}\:\mathrm{g\:s^{-1}}$.
This is likely an underestimate, since at such a low accretion rate most of the
energy would be emitted as cyclotron radiation.
For the parameters of UZ For, the calculations by \citet{1996A&A...306..232W} 
(fig. 9) predict that cyclotron cooling dominates at specific accretion rates 
$\dot{m}<10\:\mathrm{g\:cm^{-2}\:s^{-1}}$.
Therefore, if most of the energy were radiated as bremsstrahlung, accretion 
would have to occur in an unreasonably small region with $f\le10^{-8}$
($f$ is the area as a fraction of the white dwarf's surface area).
For a more typical spot size of $f=5\times10^{-4}$ \citep{1996A&A...310..526R},
we estimate from fig. 9 in \citet{1996A&A...306..232W} that the cyclotron 
luminosity is $\sim\!10^2\times L_{brems}$.
The corresponding accretion rate of $10^{13}$--$10^{14}\:\mathrm{g\:s^{-1}}$
is comparable to that determined by \citet{1996A&A...310..526R} for the low 
state of UZ For.

As shown in Fig. \ref{lightcurve}, a short flare was detected simultaneously by 
the X-ray instruments and the Optical Monitor.
The flare (Fig. \ref{flare}), which lasted $\sim\!900\:$s, 
increased the X-ray flux by a factor of $\sim\!30$.
This is comparable to the flux that we estimated for the bremsstrahlung
component during the high state (see above).
The beginning of the flare at BJD(TT) 2451924.24401(2) coincides to 
within a few seconds with the eclipse egress of UZ For
(see also Section \ref{uvlc}).
This is strong evidence that the flare emission originates from the 
white dwarf in UZ For.
We will discuss the spectral properties of the flare in Section \ref{flarespec}.


\subsection{UV light curve}
\label{uvlc}

\begin{figure}
\includegraphics{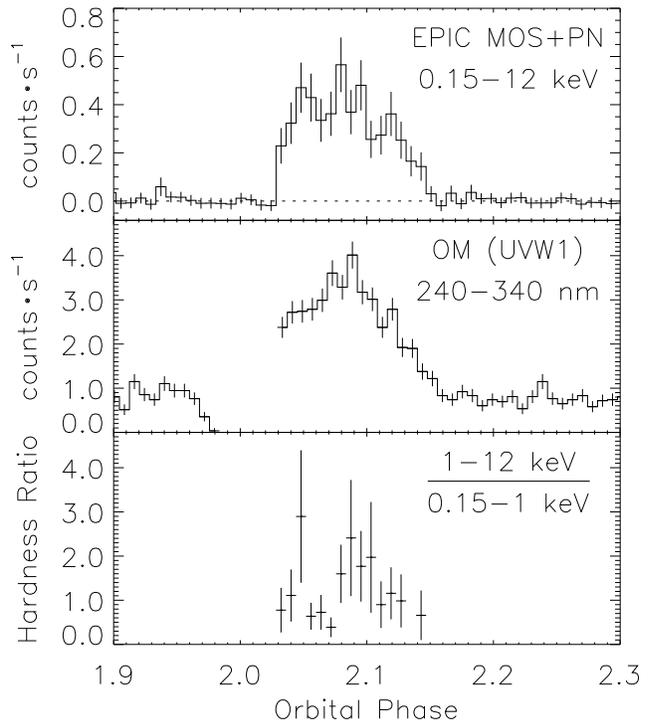}
\caption[X-ray and UV light curves of flare]
{X-ray light curve, UV light curve, and X-ray hardness ratio of the flare near 
orbital phase 2 (binning 60 s).
}
\label{flare}
\end{figure}

\begin{figure}
\includegraphics{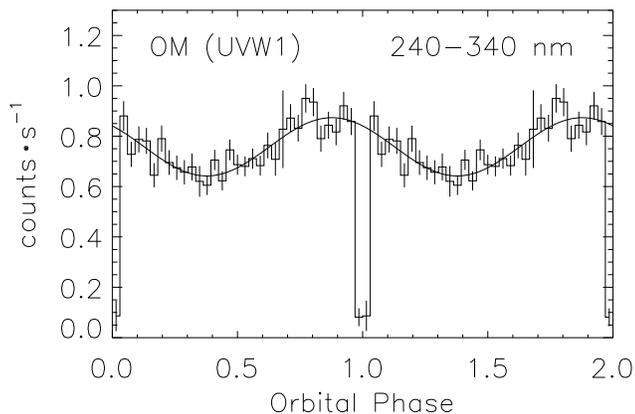}
\caption[UV modulation]
{UV light curve folded on the orbital period after exclusion of the three 
flares shown in Fig. \ref{lightcurve} (binning 230 s).
Also shown is the best fit of a sine function to the orbital modulation
(amplitude 30 per cent, maximum at phase 0.88).
}
\label{modulation}
\end{figure}

During most of the observation, the UV flux detected with the Optical 
Monitor was at a roughly constant level (Fig. \ref{lightcurve}).
Besides the flare that was also detected with the EPIC instruments, 
two more flaring events are seen near orbital phases 0 and 4.
Unfortunately, no X-ray data is available for these time periods.
It should be noted that the flaring activity is only seen inside the orbital
phase range 0.65--0.15 for which the main accretion region is visible.
This is a further indication that the flaring is caused by accretion on to 
the main pole.

We were able to identify the ingress and egress for several eclipses of 
the white dwarf by the companion star
(see Fig. \ref{lightcurve} near integer orbital phases).
Coincidentally, each of the eclipses overlapped, at least partially, with a gap 
in the OM data.
We did not detect any UV emission during the eclipses, even after integrating
all 530 s of available eclipse data.
This non-detection corresponds to an upper count rate limit of $0.05\:$s$^{-1}$ 
($3\:\sigma$) or a flux limit of
$2.5\times10^{-17}\:\mathrm{erg\:cm^{-2}\:s^{-1}}$\AA$^{-1}$ at 270 nm.
Besides the accretion stream, the only potential source of emission during 
eclipse is the M-dwarf companion.
Consistent with our limit, the models of \citet{1996A&A...305..209H}
predict a flux of
$10^{-18}$--$10^{-16}\:\mathrm{erg\:cm^{-2}\:s^{-1}}$\AA$^{-1}$
for the companion in UZ For.
Because of the fast rise and decline during eclipse,
the white dwarf is probably the sole source of the UV emission.
Due to the low accretion rate, a significant 
contribution from the accretion stream is unlikely.

The roughly constant UV count rate of $0.75\:$s$^{-1}$ that is seen during most 
of the observation is probably thermal emission from the white dwarf's 
photosphere.
The corresponding flux of
$3.8\times10^{-16}\:\mathrm{erg\:cm^{-2}\:s^{-1}}$\AA$^{-1}$ at 270 nm
is consistent with blackbody radiation from a 11000--K white dwarf
($R_{WD}=8\times10^8\:$cm, $d=200\:$pc).
The temperature is in agreement with that measured by 
\citet{1991MNRAS.253...27B}.
We find that the UV light curve (excluding the flares)
has a sinusoidal orbital modulation of $\sim\!30$ per cent
(Fig. \ref{modulation}).
The peak of the modulation is seen at orbital phase 0.88(2).
Based on the spot longitudes determined by \citet{2001MNRAS.324..899P},
we predict that emission from the main accretion region peaks at phase 
$\sim\!0.88$, while that from the secondary region peaks at $\sim\!0.03$.
This strongly suggests that the modulated flux is thermal radiation 
from the heated photosphere near the main accretion spot.
Another possible origin of the modulation is cyclotron radiation
from the accretion column.
Then, however, one would expect to see a double-peaked light curve with
peaks near orbital phases 0.15 and 0.65.
Although cyclotron radiation probably dominates the energy output from the
accretion column, only a small fraction of it ($<\!1$ per cent) is emitted
in the UV.
Assuming that the modulation is due to blackbody radiation from the main 
accretion region, we find that the flux excess of 
$1.2\times10^{-16}\:\mathrm{erg\:cm^{-2}\:s^{-1}}$\AA$^{-1}$ at 270 nm
is equivalent to a luminosity of
$4.2\times10^{26}\:\mathrm{erg\:s^{-1}}$\AA$^{-1}$ ($d=200\:$pc).
Because the accretion region is optically thick and inclined with respect to the
line of sight, we applied a geometric correction of
$\pi\cos^{-1}(\delta-i)\approx9.2$
\citep[inclination $i\approx80^\circ$, colatitude 
$\delta\approx150^\circ$; from][]{2001MNRAS.324..899P}.
As discussed in Section \ref{xraylc}, we could not identify a blackbody 
component in the X-ray spectrum.
Using this non-detection, we can place an upper limit on the blackbody 
temperature.
Depending on the hydrogen column density, which is in the range
$0.25$--$1.3\times10^{20}\:$cm$^{-2}$ \citep{1988ApJ...328L..45O},
we find a limit of 7--11 eV (80,000--130,000 K).
We can also constrain the bolometric blackbody luminosity
$L_{bb}\le1\times10^{32}\:\mathrm{erg\:s^{-1}}$
and the size of the emitting region $f\ge0.001$.
The limit on $L_{bb}$ is fairly high, $\sim\!10^3$ times higher than the 
bolometric bremsstahlung luminosity $L_{brems}$.
It is therefore likely that the actual temperature of the emitting 
region is significantly lower than the upper limit.
The lowest possible blackbody luminosity consistent with the observed UV
flux is $1.5\times10^{30}\:\mathrm{erg\:s^{-1}}$ (for $kT_{bb}=1.1\:$eV).
This is still $\sim\!20$ times higher than $L_{brems}$.
A large ratio of blackbody to bremsstrahlung luminosity is found in many polars,
typically at high accretion rates.
This "soft excess" is commonly explained by blobs in the accretion stream 
that penetrate deep into the photosphere \citep{1988A&A...193..113F}.
Blackbody radiation from this process is only emitted in the fairly small 
accretion region.
In UZ For, however, the UV flux is varying sinusoidally even 
when the main accretion spot is not visible (phases 0.15--0.65).
This is only possible if the emitting region is significantly larger than 
the accretion spot and comparable in size to the white dwarf.
Apparently, we are seeing a large heated region of the white dwarf's 
photosphere around the main accretion spot.
For a covering fraction $f=0.5$, this region has to be, on average,
$\sim\!1000\:$K hotter than the other half of the white dwarf's surface.
Since $L_{bb}$ is much larger than $L_{brems}$, reprocessing of X-rays is 
unlikely to be responsible for the heating of this region.
However, the cyclotron luminosity is expected to be much larger than $L_{brems}$
and might provide sufficient energy (see Section \ref{xraylc}).
It is also possible that we are seeing the afterglow from an earlier state of
high accretion rate,
since the time-scale for cooling of a heated pole cap is on the order of tens of
days \citep{1997PhDT........28G}.

\begin{table}
\caption{
Fitted times/phases of mid-ingress and mid-egress of the eclipses 
in the UV light curve.
The times/phases are relative to the eclipse centres as predicted by the
ephemeris in \citet{2001MNRAS.324..899P}, which has an accumulated
uncertainty of 4 s.
In all but one case, the ingress/egress duration could not be fitted 
and was fixed at 40 s.
Uncertainties are given at a $1\:\sigma$ confidence level.
}
\begin{center}
\begin{tabular}{llccc}
\hline
 & & Time & Phase & Duration \\
\hline
Ingress & phase 2  & $-225\pm4\:$s & $-0.0296(5)$ & $40\:$s \\
        & phase 3  & $-220\pm5\:$s & $-0.0290(7)$ & $40\:$s \\
Egress  & phase 0  & $+220\pm4\:$s & $+0.0290(7)$ & $40\:$s \\
        & phase 2  & $+233\pm2\:$s & $+0.0307(3)$ & $3\pm3\:$s \\
        & phase 3  & $+233\pm5\:$s & $+0.0307(7)$ & $40\:$s \\
\hline
\end{tabular}
\end{center}
\label{table1}
\end{table}

We determined the times of mid-ingress and mid-egress
by fitting the profile expected for a circular emission region
to the eclipses in the UV light curve (Fig. \ref{lightcurve}).
Except for the egress near phase 2, we could not obtain
useful constraints for the duration of ingress or egress.
We therefore assumed a fixed value of 40 s,
which is the time it takes to eclipse the entire white dwarf
\citep{1991MNRAS.253...27B}.
Although not apparent in the UV light curve (Fig. \ref{flare}), the egress near 
phase 2, just before the flare, can be clearly identified.
About 10 s after the beginning of the new OM exposure,
the count rate increased quickly from 0 to $\sim\!3\:$s$^{-1}$.
This fast rise coincides with the beginning of the flare
in the X-ray light curve.
The egress duration of $3\pm3\:$s in the UV is similar to that measured in the 
optical for the main accretion region 
\citep[$\sim\!2\:$s;][]{1991MNRAS.253...27B}.
Table \ref{table1} shows the fitted times/phases of mid-ingress and 
mid-egress relative to the eclipse centres.
Using high time-resolution optical light curves of the eclipse,
\citet{2001MNRAS.324..899P} resolved two distinct accretion regions.
They found that the egress of the main region (Spot 1) occurs at phase 0.032,
while that of the secondary region (Spot 2) occurs at phase 0.027.
The beginning of the flare at phase 0.0307 is in 
good agreement with the egress of spot 1.
This result strongly suggests that the flare was caused by accretion on to the 
main region and that the onset of the flare was hidden from us by the eclipse.
In agreement with this interpretation, the end of the flare (Fig. \ref{flare})
is seen at the same orbital phase of 0.15 at which the main accretion region 
disappears behind the white dwarf.


\subsection{Flare spectrum}
\label{flarespec}

\begin{figure}
\includegraphics{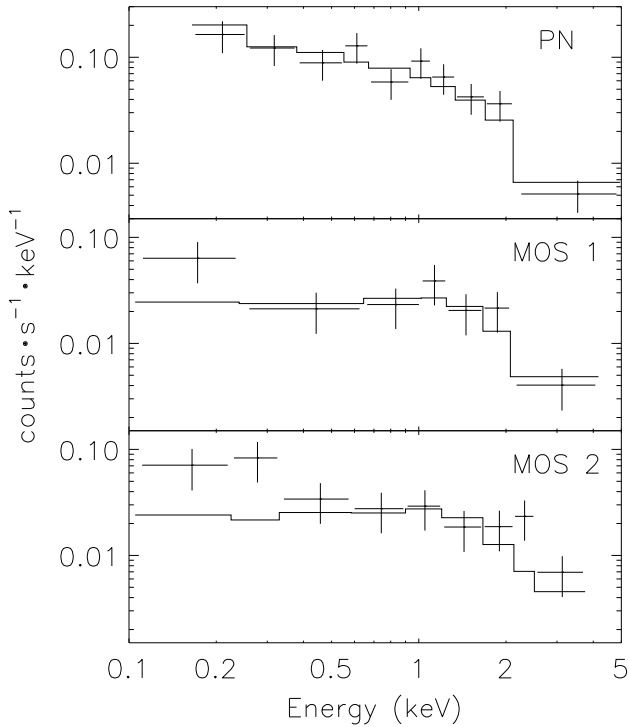}
\caption[X-ray spectrum of flare]
{Average X-ray spectrum of the flare near orbital phase 2 shown separately for 
the three EPIC detectors.
The solid lines represent the best fit of an absorbed single-temperature 
bremsstrahlung model with $kT_{brems}=6.6\:$keV 
and $N_{H}=0.5\times10^{20}\:$cm$^{-2}$.
}
\label{spectrum}
\end{figure}

As shown in the previous sections, the X-ray/UV flare near orbital phase 2
was most likely caused by a significant increase in the rate of accretion
on to the white dwarf.
According to the picture of the standard accretion column, the X-ray spectrum
of the flare should therefore consist of two spectral components,
a bremsstrahlung component from the shock-heated gas and a blackbody 
component due to reprocessing of hard X-rays in the white dwarf's atmosphere.
We find that the flare spectrum (Fig. \ref{spectrum}) is consistent with
a pure bremsstrahlung model but does not show statistically significant 
evidence for a blackbody component.
In particular, we do not find the 28 eV blackbody that dominated the 
\rosat\ spectrum during the high state \citep{1993MNRAS.262..993R},
despite the similar bremsstrahlung fluxes during both observations
(Section \ref{xraylc}).
It is plausible that, during the \xmm\ observation, accretion occurred over a 
larger area, leading to a blackbody temperature too low for a detection by the 
EPIC instruments.
A lower blackbody temperature is also expected in the absence of blobs in the 
accretion stream.
Such blobs were likely present during the \rosat\ observation, 
as \citet{1993MNRAS.262..993R} found a large soft X-ray excess.
Dense blobs impacting the white dwarf are thought to heat small areas of the 
photosphere to higher temperatures than the reprocessing of hard X-rays
\citep{1988MNRAS.235..433H}.
Without blobs in the accretion stream, the blackbody temperature would 
be much lower than the 28 eV found for the high state.
As discussed later, a blackbody component with a temperature below
$\sim\!12\:$eV was possibly detected in the UV.
The hardness ratio in Fig. \ref{flare} indicates some spectral variability,
which is likely due to irregularities in the accretion stream.

The X-ray spectrum of the flare is well fit by a single-temperature 
thermal bremsstrahlung model ($\chi_{red}^2\approx1$).
We included interstellar absorption in the model fit but could not obtain useful
constraints for the hydrogen column density $N_H$.
An upper limit of $1\times10^{20}\:$cm$^{-2}$ follows from the total 
galactic hydrogen column density in the direction of UZ For
(Dickey and Lockman 1990; Stark 1988).
\citet{1988ApJ...328L..45O} constrained $N_H$ to the range
$2.5\times10^{19}$--$1.3\times10^{20}\:$cm$^{-2}$.
Table \ref{table2} shows the fitted temperature and unabsorbed 
bolometric flux of the bremsstahlung model for various fixed column 
densities.
The temperature of $\sim\!7\:$keV is typical for 
bremsstrahlung from the accretion column and similar to that found by 
\citet{1996A&A...310..526R} from the cyclotron spectrum.
For a distance $d=200\:$pc we obtain an unabsorbed bolometric luminosity
$L_{brems}\approx2.4\times10^{30}\:\mathrm{erg\:s^{-1}}$
(we used a geometric factor of $3.1\:\pi$ as in Section \ref{xraylc}).
This luminosity corresponds to a mass accretion rate 
$\dot{M}\approx2\times10^{13}\:\mathrm{g\:s^{-1}}$
($M_{WD}=0.7\:M_\odot$, $R_{WD}=8\times10^8\:$cm).
A constant temperature model likely oversimplifies the flare spectrum,
which shows considerable temporal variability
(see hardness ratio in Fig. \ref{flare}).
However, the low signal-to-noise ratio of the data did not allow us to
obtain useful estimates for the temperature variations during the flare.
Since $\sim\!80$ per cent of the flux in the constant temperature model is
inside the 0.15--12 keV energy range and directly detected by \xmm, a
constant temperature fit should provide a good estimate for the bolometric
bremsstrahlung flux, despite the spectral variability.

\begin{table}
\caption{
Temperature and unabsorbed bolometric flux obtained from a fit of a
bremsstrahlung model to the EPIC spectra of the flare.
Results are shown for various fixed column densities $N_H$.
The uncertainties of $kT_{brems}$ are given at a $1\:\sigma$ confidence 
level.
}
\begin{center}
\begin{tabular}{ccc}
\hline
$N_H$ [cm$^{-2}$] & $kT_{brems}$ [keV]
& $F_{brems}$ [erg$\ $cm$^{-2}$s$^{-1}$] \\ \hline
$0.2\times10^{20}$ & $7.5^{+6.8}_{-2.7}$ & $6.8\times10^{-13}$ \\
$0.5\times10^{20}$ & $6.6^{+5.1}_{-2.2}$ & $6.5\times10^{-13}$ \\
$1.0\times10^{20}$ & $5.7^{+3.8}_{-1.8}$ & $6.0\times10^{-13}$ \\
\hline
\end{tabular}
\end{center}
\label{table2}
\end{table}

In the Optical Monitor, the flare (Fig. \ref{flare}) caused an average count 
rate increase of $\sim\!2\:$s$^{-1}$, which corresponds to a flux of
$1\times10^{-15}\:\mathrm{erg\:cm^{-2}\:s^{-1}}$\AA$^{-1}$ at $270\:$nm
or an integrated flux in the 240--340 nm band of the UVW1 filter
of $F_{UVW1}\approx1\times10^{-12}\:\mathrm{erg\:cm^{-2}\:s^{-1}}$.
The latter value is comparable to the bolometric bremsstrahlung flux 
$F_{brems}$.
The likely origin of the UV emission is either cyclotron radiation from the 
post-shock region or blackbody radiation from the photosphere below the shock 
(or a combination of the two).
The \xmm\ data do not allow us to distinguish between the two possibilities.

UZ For is known to be a strong emitter of cyclotron radiation
\citep{1990A&A...230..120S}.
Past observations of the cyclotron spectrum covered only optical wavelengths,
but model calculations by \citet{1996A&A...310..526R} suggest that cyclotron 
emission lines may be present in the near UV.
If the Optical Monitor did indeed detect cyclotron radiation, 
only a small fraction of the total cyclotron flux was actually seen.
Using the results by \citet{1996A&A...306..232W} in a similar way as in 
Section \ref{xraylc}, we estimate that the cyclotron luminosity was 
$10^1$--$10^2$ times higher than $L_{brems}$.
This luminosity corresponds to an accretion rate of
$10^{14}$--$10^{15}\:\mathrm{g\:s^{-1}}$ or a total accreted mass of 
$10^{17}$--$10^{18}\:$g for the duration of the flare.

The accretion region is heated by reprocessing of X-rays from the post-shock 
region or by blobs in the accretion stream that penetrate several scale 
heights into the atmosphere.
This gives rise to blackbody radiation visible at X-ray and UV energies
\citep{1995cvs..book.....W}.
Assuming that all of the UV flare emission is due to this blackbody,
we estimate a luminosity of
$4\times10^{27}\:\mathrm{erg\:s^{-1}}$\AA$^{-1}$ at 270 nm ($d=200\:$pc).
As in Section \ref{uvlc} we used a geometric factor of 9.2.
Since the blackbody component is not visible in the X-ray spectrum,
we can place an upper limit of 7--12 eV on its temperature.
The uncertainty derives mostly from the not well known hydrogen column density 
$N_{H}=0.25$--$1.3\times10^{20}\:$cm$^{-2}$ \citep{1988ApJ...328L..45O}.
(The EPIC MOS spectra in Fig. \ref{spectrum} show a small excess over the 
bremsstrahlung model at low energies.
This might be considered marginal evidence for the blackbody component
but could also be due to calibration uncertainties below 0.2 keV.)
The limit on the blackbody temperature also puts constraints on the 
effective emitting area $f_{eff}\ge0.01$ and the blackbody luminosity
$L_{bb}\le0.3-1.5\times10^{33}\:\mathrm{erg\:s^{-1}}$.
The actual luminosity is likely close to this limit, since a significantly 
smaller $L_{bb}$ would require an unreasonably large emitting area $f\gg0.01$.
In particular, it is not possible that $L_{bb}\approx L_{brems}$,
which rules out reprocessing of X-rays as a major contributor to the UV flux.
The large ratio $L_{bb}/L_{brems}\sim\!10^2$ could be due to blobs in the 
accretion stream \citep{1988A&A...193..113F}.
Then, however, one might expect $f_{eff}$ to be smaller than our limit of 0.01.
An effective emitting area $f_{eff}<10^{-4}$ is required for the 28 eV 
blackbody found by \citet{1993MNRAS.262..993R} in the \rosat\ spectrum.
(As discussed in \citet{1988MNRAS.235..433H}, $f_{eff}$ can be substantially
smaller than the size of the accretion region.)
Considering these difficulties, it seems unlikely that a large fraction of 
the UV flare emission is blackbody radiation from the accretion region.


\section{Discussion}

UZ For was found in a peculiar state that could have been classified as a 
regular low state if \xmm\ had not detected several X-rays/UV flares.
Whereas flaring is common during the high state and probably due to 
dense blobs in the accretion stream \citep{1988MNRAS.235..433H},
flaring during a low state has only been reported for two other polars.
\euve\ detected two transient events, each lasting $\sim\!1\:$hr, during a low 
state of QS Tel \citep{1993ApJ...414L..69W}.
Three possible explanations for the transients were suggested:
magnetic flares on the secondary;
dense filaments in the accretion stream impacting the white dwarf;
intermittently enhanced mass transfer from the secondary caused by magnetic
flares or coronal mass ejections.
In AM Her, \citet{shakhovskoy1993} observed an optical transient that 
lasted $\sim\!20\:$min and increased the brightness by 2 mag.
The properties of the transient (exponential decay, blue color, no polarization)
strongly suggest emission from a stellar flare on the companion star.
Also in AM Her, \citet{2000A&A...354.1003B} observed several smaller transients 
($\sim\!0.5\:$mag) that were red in color and circularly polarized.
These transients were likely caused by irregular accretion on to the white dwarf
(e.g. by filaments in the accretion stream).

Our analysis of the \xmm\ data clearly shows that the flaring in UZ For was 
caused by accretion on to the white dwarf.
This rules out magnetic flares on the secondary as the source of the X-ray and 
UV emission.
It is also unlikely that dense filaments in an otherwise constant accretion stream
caused the large increase in X-ray luminosity.
The observed bremsstrahlung spectrum and the absence of a blackbody component 
indicate that the shock was not buried in the photosphere, as is expected for 
dense filaments impacting the white dwarf.
The most likely explanation for the flaring in UZ For is an intermittent 
increase in the mass transfer rate from the companion star.

The causes of high and low states in polars are still poorly understood.
Unlike other CVs, polars do not have an accretion disk, so that changes in the 
rate of accretion on to the white dwarf must be due to variations of the mass 
flow rate through the L1-point.
Variations in the size of the secondary or of the Roche-lobe cannot be the cause 
of low states, as those changes occur on time-scales of $10^4$--$10^5\:$yr
\citep{1995ApJ...444L..37K,1988A&A...202...93R}.
However, \citet{2000ApJ...530..904H} showed that for secondaries in short-period
CVs the critical Roche-surface is well above the photosphere, so that
mass flow rates trough the L1-point are controlled by the chromosphere.
Changes in an active chromosphere can occur on time-scale short enough to 
explain the low states in polars.
Alternatively, it has been suggested that star spots may appear at the 
L1-point and greatly reduce the mass transfer rate
\citep{1994ApJ...427..956L,1998ApJ...499..348K,2000A&A...361..952H}.
However, \citet{2000ApJ...530..904H} point out that, for short-period CVs, star 
spots exist well below the critical Roche surface, and the magnetic 
activity near the spots might actually increase the accretion rate.
In both scenarios, it is possible that coronal mass ejections or solar flares 
near the L1-point intermittently increase the mass transfer rate,
thus causing accretion events as those seen in UZ For.

Little is known about the level of stellar activity in CV secondaries.
But because of their rapid rotation, which is synchronized with the binary 
orbital motion, an active atmosphere can be expected.
A number of observations suggest that star spots and magnetic flares do exist
on CV secondaries
\citep[e.g.][]{2000ApJ...530..904H,2002ApJ...568L..45W,shakhovskoy1993}.
We showed that the X-ray transient in UZ For was caused by accretion of 
$10^{17}$--$10^{18}$g of gas on to the white dwarf.
It is not unreasonable that this much mass was ejected by a flare on the 
dM4.5 companion star.
From observations of stellar flares on nearby M dwarfs, 
\citet{1990A&A...228..403P} estimated a density range of 
$\le\!10^{12}$--$10^{13}\:$cm$^{-3}$ and a volume range of 
$\ge\!10^{27}$--$10^{29}\:$cm$^3$, corresponding to flare masses of 
$10^{15}$--$10^{18}\:$g.
For the largest flares, peak X-ray luminosities up to 
$\sim\!10^{30}\:\mathrm{erg\:s^{-1}}$ (0.05--2 keV) were measured.
A flare with such a high luminosity would have been detected by \xmm\
with an EPIC count rate of $\sim\!0.2\:$s$^{-1}$ over $\sim\!10^3\:$s.
Since no emission from the companion star was seen, the X-ray luminosity
of any stellar flare must have been below
$\sim\!\:10^{29}\:\mathrm{erg\:s^{-1}}$.
Coronal X-ray emission other than that from large stellar flares is
unlikely to be detectable, even with the data integrated over the
entire observation.
For the most active stars of type dM5 or later,
\citet{1993ApJ...410..387F} measured average X-ray luminosities up to
$10^{28}\:\mathrm{erg\:s^{-1}}$ (0.1--2.4 keV).
The corresponding EPIC count rate of $\sim\!0.002\:$s$^{-1}$ is slightly
below the detection threshold for the UZ For observation.
We conclude that, despite the absence of X-ray emission from the companion
star, accretion of gas ejected by stellar flares is a viable explanation
for the transient events observed in UZ For.


\section{Conclusion}

During the \xmm\ observation, UZ For was found in an extremely low accretion 
state with an X-ray luminosity $\sim\!800$ times fainter than during a high 
state previously observed with \rosat.
Occasional X-ray and UV flaring was detected by the X-ray instruments and the 
Optical Monitor.
The largest flare lasted $\sim\!900\:$s and increased the X-ray flux by a 
factor of $\sim\!30$.
We found that the beginning of this flare coincided to within a 
few seconds with the eclipse egress of the main accretion region.
This provides strong evidence that the flaring was caused by accretion on to the 
white dwarf.
The X-ray spectrum of the flare is consistent with $\sim\!7\:$keV thermal 
bremsstrahlung from the accretion column.
A blackbody component, as seen with \rosat\ during the high state, was not 
found.
It is plausible that, because of a larger accretion region or the absence of 
blobs in the accretion stream, the blackbody temperature was too low for a
detection by the X-ray instruments.
The increase in the UV flux seen during the flare was probably caused by 
cyclotron radiation from the accretion column.
A significant contribution of blackbody radiation to the UV flare emission is 
unlikely as this would require a very large soft excess.
Under the assumption that all accretion energy is emitted as bremsstrahlung, we 
estimate an accretion rate of $2\times10^{13}\:\mathrm{g\:s^{-1}}$ during the 
flare.
However, since at this low rate most of the energy is emitted as cyclotron 
radiation, the actual accretion rate was probably $10^1$--$10^2$ times higher.
We therefore estimate that during the flare a total of $10^{17}$--$10^{18}\:$g 
of gas was accreted on to the white dwarf.
The likely cause of the flaring observed in UZ For is stellar activity 
on the companion star that intermittently increased the mass transfer rate 
near the L1-point.
The mass that was accreted on to the white dwarf during the large transient is 
consistent with the mass ejected by a stellar flare.

Before and after the X-ray transient, extremely weak X-ray emission, possibly 
due to the regular low-state accretion on to the main region, was detected.
The observed X-ray luminosity corresponds to an accretion rate of
$6\times10^{11}\:\mathrm{g\:s^{-1}}$.
Since cyclotron radiation dominated the energy output,
the actual accretion rate was probably $10^{13}$--$10^{14}\:\mathrm{g\:s^{-1}}$,
which is similar to the rates estimated for previous low states.
In addition to the flare emission, we detect a roughly constant UV flux 
consistent with blackbody radiation from a 11000--K white dwarf.
A small orbital modulation of the UV flux indicates the presence of a large, 
heated pole cap around the main accretion region.

Flaring during a low state has only been observed for a few polars.
This may be due to insufficient monitoring of polars in low states.
Yet low-state observations are essential since, during high and intermediate 
states, flaring caused by stellar activity on the companion star is likely 
overlooked and mistakenly attributed to accretion stream instabilities.
Future low-state observations will reveal how common flaring due to stellar 
activity is among polars.
An interesting question that might also be answered is,
whether flaring occurs preferentially at the beginning or end of low states.
Monitoring of irregular accretion in low-state polars may provide a new way to 
study stellar flares or other types of mass ejections that are too faint to be 
observed directly.


\section*{acknowledgements}

This work is based on observations obtained with \xmm,
an ESA science mission with instruments and contributions
directly funded by ESA Member States and the USA (NASA).
The authors acknowledge support from NASA grant NAG5-7714.


\bibliographystyle{mn2e}
\bibliography{uzfor}

\label{lastpage}

\end{document}